\documentclass[12pt,preprint]{aastex}

\usepackage{graphicx,color,soul}

\shorttitle{SO isotopologues}
\shortauthors{Lattanzi et al.}

\begin{document}

\title{Rare isotopic species of sulphur monoxide:\\ the rotational spectrum in the THz region}

\author{Valerio Lattanzi} \affil{Dipartimento di Chimica \lq\lq Giacomo Ciamician'', Universit\`{a} di Bologna, Via Selmi 2, I-40126 Bologna, Italy; \\  Max-Planck-Institut f\"ur Extraterrestrische Physik, Giessenbachstra{\ss}e 1, 85748 Garching, Germany} \email{lattanzi@mpe.mpg.de}
\and \author{Gabriele Cazzoli, Cristina Puzzarini} \affil{Dipartimento di Chimica \lq\lq Giacomo Ciamician'', Universit\`{a} di Bologna, Via Selmi 2, I-40126 Bologna, Italy}

\begin{abstract}

Many sulphur-bearing species have been detected in different astronomical environments and have allowed to derive important information about the chemical and physical composition of interstellar regions. In particular, these species have also been showed to trace and probe hot-core environment time evolution. Among the most prominent sulphur-bearing molecules, SO, sulphur monoxide radical, is one of the more ubiquitous and abundant, observed also in its isotopic substituted species such as $^{34}$SO and S$^{18}$O. Due to the importance of this simple diatomic system and to face the challenge of modern radioastronomical facilities, an extension to THz range of the rare isotopologues of sulphur monoxide has been performed. High-resolution rotational molecular spectroscopy has been employed to extend the available dataset of four isotopic species, SO, $^{34}$SO, S$^{17}$O, and S$^{18}$O up to the 1.5~THz region. The frequency coverage and the spectral resolution of our measurements allowed a better constraint of the molecular constants of the four species considered, focusing especially for the two oxygen substituted isotopologues. Our measurements were also employed in an isotopically invariant fit including all available pure rotational and ro-vibrational transitions for all SO isotopologues, thus enabling accurate predictions for rotational transitions at higher frequencies. Comparison with recent works performed on the same system are also provided, showing the quality of our experiment and the improvement of the datasets for all the species here considered. Transition frequencies for this system can now be used with confidence by the astronomical community well into the THz spectral region.
\end{abstract}


\section{Introduction}

Molecules with sulphur account for about 10\% of the species identified in the interstellar gas and circumstellar envelopes, and among them the most prominent is sulphur monoxide (SO). Since its first detection by \citet{cag73} towards Orion A, many astronomical studies have been dedicated to this molecule, including those on the gas kinematics \citep{1982ApJ...259..617P}, molecular abundances \citep{1987ApJ...315..621B}, and spatial distribution \citep{1995ApJS...97..455S}. Sulphur-bearing species are considered good probes of hot-core time evolution, since they are highly sensitive to physical and chemical variations of these regions \citep{2001A&A...370.1017V, 1998A&A...338..713H}. Thus, they can be used as tools for investigating the chemistry and physical properties of complex star-forming regions (SFRs) located in dense molecular clouds. On the other hand, it is known that some molecules (SiO, H$_2$CS, SO, SO$_2$) show increased abundances in regions affected by shocks \citep{1996ARA&A..34..111B} as a result of the action of outflows on the surrounding gas. The study of molecular lines from shocked areas provides valuable information about the chemical processes and the physical conditions of the shocked components.

Recently \citet{2013A&A...556A.143E} dedicated an entire study to the sulphur oxide species in Orion KL, detecting 68 lines of SO, $^{34}$SO, $^{33}$SO and S$^{18}$O and providing upper limits for abundance of rarer isotopologues, such as S$^{17}$O, $^{36}$SO, and $^{34}$S$^{18}$O. Making use of the IRAM-30m antenna and analysing the SO non-LTE excitation they found that the most of the emission of SO arises from the High Velocity Plateau (column density $N=(5\pm1)\times 10^{16}$cm$^{-2}$), and the Hot Core. These results confirm how SO (and SO$_2$) is a good tracer not only of shock-affected areas, but also of hot dense gas. The analysis of the abundance ratio of the SO isotopologues compared to that of SO$_2$ for the different component of the observed region, showed that the latter is five time more abundant than SO in the High Velocity Plateau and the Hot Core. This work clearly elucidated the importance of isotopic analysis in the interstellar medium to derive information both on the abundance of optically thick parent species, and on the evolution of the isotopic ratio in different astronomical environments.

More recently, a new extensive laboratory spectroscopic characterisation of sulphur monoxide isotopologues has been published by Martin-Drumel et al. (2015, hereafter MD). In this work the authors provided new high-frequency measurements for the main isotopic species and its sulphur isotopic substituted ($^{34}$SO and $^{33}$SO) using continuous-wave terahertz photomixing based on a frequency comb, with a frequency accuracy ranging from 10 to 800 kHz. Furthermore, the authors combined their measurements with all available pure rotational and ro-vibrational transitions of different SO isotopologues in an isotopically-invariant fit.

In the work here presented, we report the extension to the THz-range of the available dataset for sulphur monoxide and the minor isotopologues $^{34}$SO (4.2\%), S$^{18}$O (0.2\%) and S$^{17}$O (0.04\%) through laboratory absorption spectroscopy, in order to improve the available spectroscopic parameters and determine centrifugal distortion terms. The determination of the latter has also supported by high-level quantum-chemical calculations. Finally, a detailed comparison with previous available spectroscopic data is supplied in the discussion section, together with the results from an isotopic invariant fit involving all available rotational and ro-vibrational frequencies for all isotopologues of SO.\\

\section{Experiment}

\begin{figure}
\resizebox{\hsize}{!}{\includegraphics[width=\textwidth]{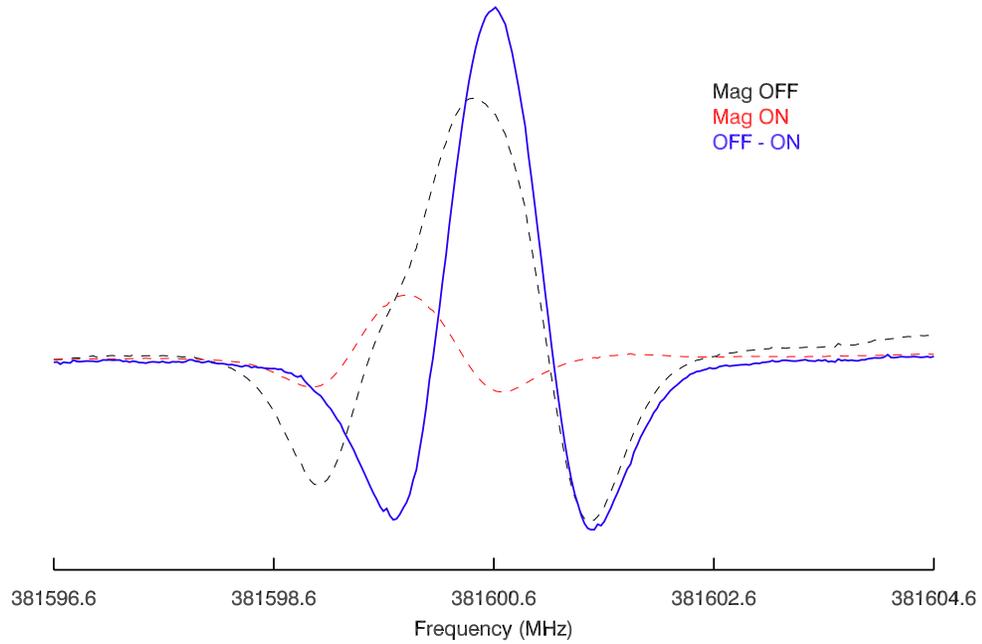}} \caption{The $N,J$ =
9,10 $\leftarrow$ 8,9 rotational transition of $^{34}$SO. The
black-dashed shows the first acquired spectrum, in which a
contaminant species transition is on the left edge of the $^{34}$SO
rotational transition. The effect of applying a magnetic field is to
cancel the radical transition, isolating the non-magnetic
contaminant transition (in red), whose contribution can then be easily
subtracted to derive a precise information on the radical of
interest (in blue). Owing to the modulation and detection scheme
employed, the instrumental lineshape is approximately the second
derivative of a Lorentzian. No averaging time was required for
obtaining a good S/N.} \label{fig:1}
\end{figure}

All the measurements have been carried out with a frequency-modulated, computer-controlled spectrometer (65 GHz - 1.6 THz) \citep{THZ1, hdo17, LD-THz}, with the 200 GHz -- 1.5 THz range actually considered. Frequency multipliers driven by Gunn diode oscillators, phase-locked to a rubidium frequency standard, have been employed as the millimeter and sub-millimeter-wave sources. The frequency modulation was obtained by sine-wave modulating the 72 MHz local oscillator of the synchronisation loop at 16.66~kHz, with the modulation depth varied from 400~kHz to 2.4~MHz according to the transition frequency under consideration. Liquid-He-cooled InSb and Schottky diode detectors were used and their output processed by means of a lock-in amplifier tuned to twice the modulation frequency (i.e., second harmonic detection was performed). Previous unrelated experiments performed in the cell with sulphur-bearing chemicals allowed us to achieve very good signals of sulphur monoxide only with the flow of $\sim$~30~mTorr of O$_2$ and a DC discharge of 12~mA, at room temperature. The measurements of the $^{34}$S isotopic species have been carried out in natural abundance (4.22\%), while an isotopically enriched oxygen sample, a mixture of 10\% $^{17}$O$_2$ and 90\% $^{18}$O$_2$, has been employed for the experiment of the two oxygen minor isotopologues. A few-Gauss magnetic field, parallel to the radiation source, has been used to disentangle the free-radical signal from those of closed-shell species whose transitions contaminated the spectral window. Subtraction of the on-field spectrum to the off-one allowed a perfect zeroing of the effect of the undesired non-magnetic species (see Fig.~\ref{fig:1}).\\

\section{Computations}

As mentioned in the Introduction, to guide the analysis and the experimental determination of the sextic centrifugal distortion constant, quantum-chemical calculations at the coupled-cluster (CC) singles and doubles (CCSD) approach augmented by a perturbative treatment of triple excitations \citep[CCSD(T);][]{1989CPL...157..479R} were performed with the cc-pCVQZ basis set \citep{1989JChPh..90.1007D, 1995JChPh.103.4572W, 2002JChPh.11710548P}. All electrons were included in the correlation treatment.

For each isotopic species considered (SO, $^{34}$SO, S$^{17}$O, S$^{18}$O), the harmonic force field was obtained using analytic second derivatives \citep{ccsec}, whereas the cubic force field was determined in a normal-coordinate representation via numerical differentiation of the harmonic force constants \citep{thiel2, SLG98}. The quartic and sextic centrifugal-distortion constants were then determined by means of vibrational perturbation theory \citep{mills, re-ch2fi}.

All computations were performed with the {\sc CFour} program package \citeyearpar{cfour}.\\

\section{Analysis}

The total angular momentum \textbf{J} for a diatomic molecule in the $^3\Sigma$ state is given by the coupling between the molecular angular momentum \textbf{N} and the electronic spin momentum \textbf{S}. The effective Hamiltonian operator may be expressed as the sum of three terms: $$H = H_{rot}+H_{S-N}+H_{S-S}$$ with $$H_{rot} = \mathbf{N}^2(B-D\mathbf{N}^2+H\mathbf{N}^4)$$\\ being the rotational Hamiltonian. For each \emph{N} value (being $N$ the associated quantum number to the angular momentum operator), there are three different energy levels corresponding to the three possible space orientation of the \textbf{S} vector (\emph{S}=0,$\pm$1). $$H_{S-N}=\mathbf{N}\cdot\mathbf{S}(\gamma_B +\gamma_D\mathbf{N}^2)$$\\ and $$H_{S-S}=\frac{2}{3}(\lambda_B +\lambda_D\mathbf{N}^2)(3S_z^2-\mathbf{S}^2)$$ are Hamiltonians describing the electron spin-rotation and spin-spin interactions, respectively. $B, \gamma_B$, and $\lambda_B$ are the rotational, the electron spin-rotation and the spin-spin interaction constants, and $D,H,\gamma_D$, and $\lambda_D$ represent their centrifugal distortion correction. The derived parameters of the Hamiltonian for each isotopic species are summarised in Table~\ref{tab:1}-\ref{tab:2}.

The effective Hamiltonian of the $^{17}$O species also includes the nuclear hyperfine interactions with the overall rotation, due to the $^{17}$O nuclear spin  ($I=5/2$): $$H_{hfs}=b_F\mathbf{I\cdot S}+c(I_zS_z-\frac{1}{3}\mathbf{I\cdot S})+eQq\frac{(3I_z^2-\mathbf{I}^2)}{4I(2I-1)}+C_I\mathbf{I\cdot N}$$ where $b_F$ and $c$ are the isotropic (Fermi contact interaction) and anisotropic parts of the electron spin-nuclear spin coupling, $eQq$ is the nuclear quadrupole coupling constant, and $C_I$ is the nuclear spin-rotation interaction constant.

Sulphur monoxide has been the subject of several laboratory studies. The Cologne Database for Molecular Spectroscopy \citep{2005JMoSt.742..215M} has been used as a review of the previous works, and to obtain a starting dataset to flag transitions to remeasure with higher accuracy or to derive higher frequency predictions. The first spectroscopic study for the fundamental isotopic species date back to the late 70s \citep{1976JMoSp..60..332C}, followed in the years by many other works, including some THz measurements \citep{1994JMoSp.167..468C}. Also for the $^{34}$SO previous laboratory data cover quite a large range of the mm/sub-mm wave spectrum, with some measurements taken at 1.05~THz by \citep{1996JMoSp.180..197K}, allowing a total of 42 rotational transition of the $X^3\Sigma^-$ electronic ground state catalogued. Somewhat more critical have been in the past years the measurements for the minor oxygen isotopic substituted species, such as S$^{18}$O and S$^{17}$O. Both species have been the subject of the same work of the $^{34}$S species, but while for the former the measurements have been carried out up to the THz region (1.03~THz), the rarer $^{17}$O species has been studied up to 620~GHz. In all the cases mentioned above, the accuracy of the measurements spans from 50~kHz, in the lower frequency range, up to to 100~kHz in the THz region, values that our measurements can easily improve.

As briefly mentioned in the introduction, during the conclusion of the present study, a new spectroscopic analysis on the SO isotopologue system has been published \citep{2015ApJ...799..115M}. The focus of this work has been the extension of the absorption spectrum of SO, $^{34}$SO, and $^{33}$SO up to 2.5~THz, allowing to perform a global isotopically-invariant fit.
In the present work, for the main isotopic species, two rotational transitions in the 340~GHz region have been remeasured and 10 new transitions have been detected in the 1.07--1.41~THz frequency range. The regions 200--500~GHz and 1.10--1.40~THz have been studied for detecting 23 and 24 rotational transitions of the $^{34}$SO and S$^{18}$O isotopic species, respectively. Finally, the rarest isotopologue S$^{17}$O has been observed through 15 rotational transitions in the 1.10--1.50~THz region. The accuracy of our measurements is 20-30 kHz in the millimeter-/submillimeter-wave range and 50 kHz in the THz region. The measured frequencies will be made available for the spectroscopic community through the CDMS on line catalogue.

\section{Discussion}

The results of these new measurements are summarised in Tables \ref{tab:1} and \ref{tab:2}, for SO and $^{34}$SO (rms~=~0.62 and 0.86), and S$^{17}$O and S$^{18}$O (rms~=~0.74 and 0.54), respectively, and compared with the data obtained by MD. From this comparison we observe that our laboratory data for SO and $^{34}$SO are in a very good agreement with the corresponding constants derived from an effective-fit for the same species (see Table 2 of MD). The higher frequency measurements performed in the previous work for the main species allowed the authors to constrain the second-order centrifugal distortion correction of the spin-spin interaction constant, $\lambda_2$, undetermined in our analysis. All other terms show an agreement of better than 1~\%, except for the sextic centrifugal distortion parameter, $H$: in this case our parameters are larger than those derived previously by 11 and 19\%, respectively for SO and $^{34}$SO. To inspect this large deviation, we make use of our computed values and of an empirical, well-tested scaling procedure: $$H^{iso}_{scal} = H^{iso}_{calc} \times (H^{main}_{exp} / H^{main}_{calc}) $$ \\ where the superscripts $iso$ and $main$ refer to a specific isotopic species and the main isotopologue, respectively; $scal$, $exp$, and $calc$ denote the scaled, experimental, and quantum-chemically calculated values for $H$, respectively. Using the value of -6.974(22)$\times$10$^{-9}$ MHz for SO, we obtain a scaled value of -6.56$\times$10$^{-9}$ MHz for $^{34}$SO, which is roughly in between our and MD values and well within the confidence range of our datum. This outcome suggests that the fit of MD might take some benefit from the inclusion of the present measurements. In Table \ref{tab:2} the results for the two oxygen-substituted species are collected. Here, the literature values reported for the comparison are those predicted from the isotopically-invariant fit available in MD. As for the previous table, the agreement for all the parameters is better than few percent. The only relevant deviation is noted for the $H$ constant of the S$^{17}$O species: -6.31(28)$\times$10$^{-9}$ MHz vs -5.74$\times$10$^{-9}$ MHz. Once again, we resort to the scaled computed value (-6.18$\times$10$^{-9}$ MHz) to a deeper understanding of such a discrepancy. The latter value well agrees, within the given uncertainty, with our datum, while it suggests that the $H$ constant from the isotopically-invariant fit is underestimated. Therefore, this fit would largely benefit by the inclusion of our recorded frequencies. Since the Dunham expansion model MD used to determine the predicted parameter does not include hyperfine interactions, in Table~\ref{tab:2} the corresponding parameters are compared with the results from \citet{1996JMoSp.180..197K}. A very good agreement is observed. The only comment concerns the $^{17}$O spin-rotation constant, which is badly determined by the fit: we decided to keep it fixed at the quantum-chemical value, computed at a level of theory known to provide quantitative agreement with experiment \citep{2010IRPC...29..273P}. For S$^{18}$O, we notice that the scaled value of $H$ is -5.53$\times$10$^{-9}$ MHz and both our and MD values seem to be underestimated. On the other hand, our datum is affected by a large uncertainty and the scaled value is within the upper limit.

\begin{figure}
\resizebox{\hsize}{!}{\includegraphics{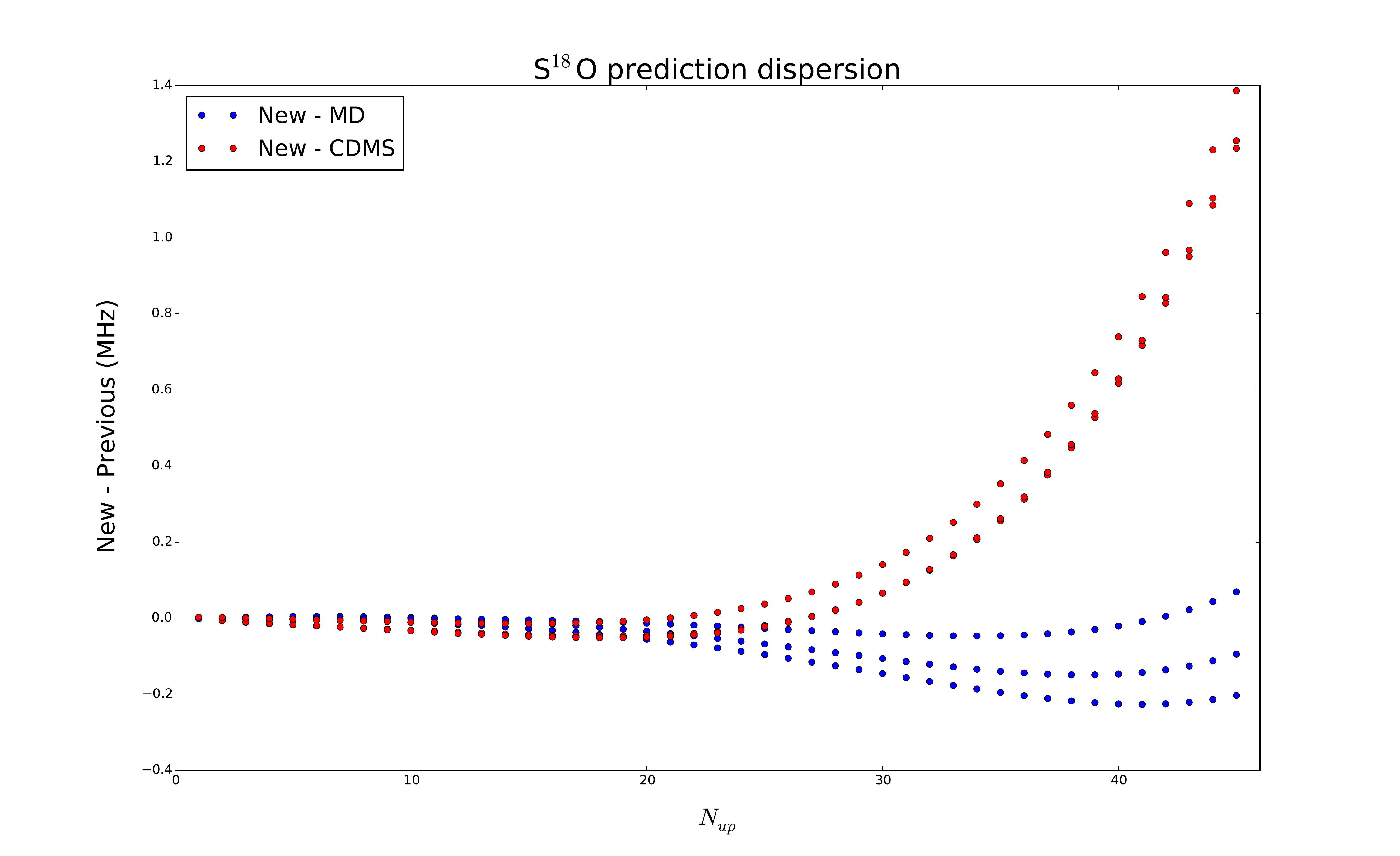}}
\resizebox{\hsize}{!}{\includegraphics{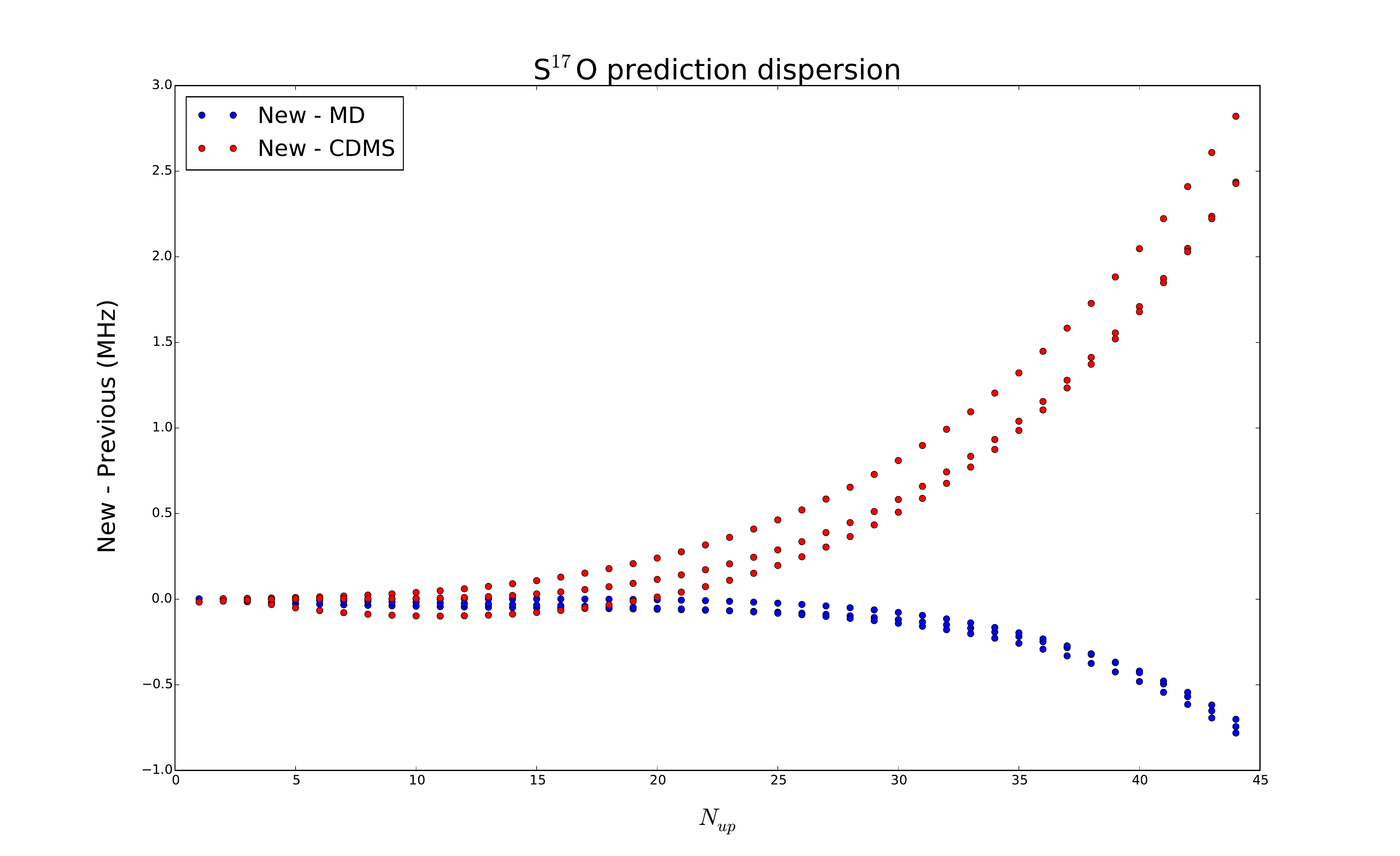}}
\caption{Comparison of the predictions ($<$~2~THz) of the $\Delta N,
\Delta J =+1$ (with  $J=N, N\pm1$) transitions, for the two oxygen
isotopic substituted species. For each $N$ the three $J$ levels are
shown. \lq\lq New'' predictions are based on our data set.}
\label{fig:2}
\end{figure}

It is also worth to compare now, for the oxygen isotopologues of sulphur monoxide, all the three available datasets, specifically ours, the MD one, and the catalogue on the CDMS website (\underline{www.cdms.de}). In Figure \ref{fig:2} this comparison is made for the predicted transitions in the frequency range 60--2000~GHz; in particular, it is illustrated the difference of the strongest set of transitions ($\Delta N, \Delta J =+1$, with  $J=N, N\pm1$), the more relevant for the astrochemistry community, among our dataset and the two previous ones. The need of new measurements is particularly evident when comparing our new predictions with those available on the CDMS catalogue, which, in the high-frequency range ($\gtrsim$~1~THz) differ by more than 1~MHz from ours, for the $^{17}$O isotopic substituted species. Less significant but still sizeable is the comparison with the Dunham isotopically-invariant results. First, it is important pointing out the overall quality of these predictions, which make use of a very large dataset of measurements of different isotopic species and several vibrational excited states. Second, it must be noted that also in this case, although less prominent than in the previous one, the deviation at high-frequency might be worth to consider, and very likely related to the discrepancy observed for the sextic centrifugal distortion term. The two datasets show a considerable better agreement for the other oxygen species, while a larger difference can be noted comparing our predictions with the older data.

To conduct a more exhaustive comparison with the previous data and to inspect the impact of our new measurements, a multi-isotopologue Dunham analysis of the full dataset has been performed, using the known reduced mass dependences given by: $$Y_{l,m} = U_{l,m}\left(1+\frac{\Delta^S_{l,m}m_e}{M_S}+\frac{\Delta^O_{l,m}m_e}{M_O}\right)\left(\frac{1}{\mu}\right)^{\frac{l}{2}+m}$$ for the rotational and vibrational parameters; $$X_{l,m} = U^X_{l,m}\left(1+\frac{\Delta^{X,S}_{l,m}m_e}{M_S}+\frac{\Delta^{X,O}_{l,m}m_e}{M_O}\right)\left(\frac{1}{\mu}\right)^{\frac{l}{2}+m}$$ with $X = \lambda, b_F, c, eQq$; and $$X_{l,m} = U^X_{l,m}\left(1+\frac{\Delta^{X,S}_{l,m}m_e}{M_S}+\frac{\Delta^{X,O}_{l,m}m_e}{M_O}\right)\left(\frac{1}{\mu}\right)^{1+\frac{l}{2}+m}$$ with $X = \gamma, C_I$. 
The $\Delta^i_{l,m}$ and $\Delta^{X,i}_{l,m}$ are unitless coefficients which account for the effect of Born-Oppenheimer breakdown.
For a clearer relation of the Dunham's coefficients with the molecular energy levels and the rotational-vibrational parameters see Townes \& Schawlow (\citeyear{townes}) equations 1-35 and 1-36. The SPFIT/SPCAT software \citep{P91} has been used for this analysis and the corresponding parameters are shown in Table \ref{tab:3}. The SPFIT parameters have been converted to the isotopically invariant terms $U_{l,m}$ through the $Y_{l,m}$ and $X_{l,m}$ relations aforementioned, for a better comparison with the previous published data; more specifically the parameters derived from the present work are set side by side with the ro-vibrational isotopically constants of MD, along with the relative reduced rms errors. Some high-order parameters derived in MD have been left out from our analysis since their inclusion in the global analysis did not contribute to the quality of the final fit. The comparison shows a general very good agreement for all the parameters analysed. Nevertheless, there are some discrepancies that required a deeper look into. The vibrational corrections to the rotational constants, $U_{3,1}$, $U_{4,1}$, and $U_{5,1}$, show a difference from a factor few to more than 10. Surprisingly, both the former two parameters are in very good accordance (in 1$\sigma$ deviation) with the isotopically invariant analysis performed by \citet{1996JMoSp.180..197K} to which MD compared their work (see Table 4 of MD). Unfortunately this comparison can not be addressed for the third parameter $U_{5,1}$, which have been determined for the first time by MD.
A similar situation can be seen for the higher order vibrational corrections to the $\lambda$ constant, namely the $U^{\lambda}_{4,0}$, $U^{\lambda}_{5,0}$ and $U^{\lambda}_{6,0}$ parameters. The former constant is the one showing the largest deviation, including a change in sign, from -0.1057 to 0.6547 MHz$\cdot$amu$^2$. Again, as for the previous cases, a more accurate analysis reveals a good accordance (in 1$\sigma$ deviation) with the value derived by \citet{1996JMoSp.180..197K}. Similarly, the other two parameters, $U^{\lambda}_{5,0}$ and $U^{\lambda}_{6,0}$, have been determined for the first time by MD and can not be compared to a third analysis: here the difference is less critical among the two studies but there is still a factor $\sim$4 and $\sim$~10, respectively.

Crucial parameters of an isotopically invariant analysis are the Born-Oppenheimer Breakdown constants. In Table \ref{tab:4} the BOB constants derived from our work are compared with those from MD. The parameters for the correction to the rotational constants, $\Delta^{S}_{0,1}$ and $\Delta^{O}_{0,1}$, show a very good agreement with the latter slightly ($\sim\times$2) improved with our global analysis. Similar good accordance is found for the sulphur correction to the vibrational frequency, $\Delta^{S}_{1,0}$, while a large discrepancy, including a change in sign, is shown for the oxygen correction, $\Delta^{O}_{1,0}$. A future study including more ro-vibrational and vibrational transitions of oxygen substituted species (currently there is only the first excited vibrational band of S$^{18}$O in the analysis) might help to elucidate this discrepancy and further constrain this important parameter. To compare the BOB coefficients of the fundamental $\lambda$ constants, $\Delta^{\lambda,S}_{0,0}$ and $\Delta^{\lambda,O}_{0,0}$, the equation $$\Delta^A_{0,0} = -\frac{M^0_A}{m_e(U_{0,0}+X^A_{0,0}+X^B_{0,0})}X^A_{0,0}$$ has been used. Here $A$ is the atom considered, $M^0_A$ is the atomic mass of the main isotopic species of the $A$ atom, $m_e$ is the electron mass, $U$ is the Dunham constant to which the BOB coefficient is related, and the $X$'s are the parameter provided from MD analysis. The oxygen BOB coefficient is in very good agreement with that derived in the previous work, and slightly improved ($\sim\times$2); the sulphur one, $\Delta^{\lambda,S}_{0,0}$, is dissimilar and, once again, opposite in sign. As seen previously for the discussion about the $U$'s, a comparison of our analysis with that of \citet{1996JMoSp.180..197K} reveals a better accordance for this parameter, with respect to its sign (now both positive) and its absolute values, although different to few sigma deviation (0.886 and 0.221, our and Klaus et al. \citeyear{1996JMoSp.180..197K}, respectively).

The present laboratory work extends the spectroscopic study of a very simple and important diatomic radical well into the THz region, including some minor isotopic species. For such an important molecular system, widely distributed and very abundant in many different interstellar regions, the accuracy of the dataset of less abundant isotopologues is also crucial. This is particularly true in an era in which high-sensitivity, high-spectral and spatial resolution radioastronomical facilities are becoming available to the scientific community. The ALMA interferometer has become the reference and most relevant observatory for the mm- and sub-mm-wave region; its unprecedented properties will allow to observe SO and its isotopologues in regions previously undetected. Above the THz threshold the instruments of the airborne observatory SOFIA will be also able, in the near future, to possibly extend the identification of this molecular system in the interstellar environments at higher frequency.\\

\acknowledgments

This work has been supported in Bologna by `PRIN 2012' funds (project ``STAR: Spectroscopic and computational Techniques for Astrophysical and atmospheric Research'') and by the University of Bologna (RFO funds). The authors would like to thank Marie Aline Martin-Drumel for providing the manuscript before its publication.

%
%
%
%

%


\begin{deluxetable}{lr@{.}lr@{.}lrr@{.}lr@{.}l}
\tablecolumns{10} 
\tablewidth{0pc}
\tablecaption{Spectroscopic constants$^a$ of SO and $^{34}$SO (in MHz)} 
\tablehead{  \colhead{} & \multicolumn{4}{c}{SO}  &  \colhead{} & \multicolumn{4}{c}{$^{34}$SO}  \\
\cline{2-5} \cline{7-10} \\
 & \multicolumn{2}{c}{This work} & \multicolumn{2}{c}{MD$^b$} & \colhead{} & \multicolumn{2}{c}{This work} & \multicolumn{2}{c}{MD$^b$}} 
\startdata
$      B$        &     21523&55568(26)          &    21523&555878(79)     &    &    21102&73094(43)         &     21102&73228(31)         \\
$  10^3D$        &         33&91431(60)         &       33&915261(85)     &    &       32&59862(90)          &        32&60073(55)         \\
$  10^9H$        &        --7&73(37)            &      --6&974(22)        &    &      --7&22(49)            &       --6&06(31)            \\
$ \lambda_B$     &     158254&3860(80)          &   158254&392(09)        &    &   158249&715(12)           &    158249&807(24)           \\
$ \lambda_D$     &          0&30662(12)         &        0&306259(72)     &    &        0&30030(38)         &         0&30063(08)         \\
$ 10^6\lambda_H$ &  \multicolumn{2}{c}{--}      &        0&478(47)        &    &  \multicolumn{2}{c}{--}    &   \multicolumn{2}{c}{--}    \\
$  \gamma_B$     &      --168&3045(22)          &    --168&3043(15)       &    &    --164&9863(33)          &      --164&9948(36)         \\
$ 10^3\gamma_D$  &        --0&5239(46)          &      --0&52545(97)      &    &      --0&5181(61)          &        --0&5116(27)         \\
\hline
$\sigma^2$       &           0&62               &        0&88             &    &        0&86                &          0&85             \\
\enddata
\begin{itemize}
\item [$^a$] Uncertainties (1$\sigma$) are in units of the last significant digit.
\item [$^b$] Martin-Drumel et al. (\citeyear{2015ApJ...799..115M}).
\end{itemize}\label{tab:1}
\end{deluxetable}



\begin{deluxetable}{lr@{.}lr@{.}lrr@{.}lr@{.}l}
\tablecolumns{10}
\tablewidth{0pc}
\tablecaption{Spectroscopic constants$^a$ of sulfur monoxide oxygen minor isotopologues (in MHz)} 
\tablehead{  \colhead{} & \multicolumn{4}{c}{S$^{18}$O}  &  \colhead{} & \multicolumn{4}{c}{S$^{17}$O}  \\
\cline{2-5} \cline{7-10} \\
& \multicolumn{2}{c}{This work} & \multicolumn{2}{c}{MD$^b$} & \colhead{} &\multicolumn{2}{c}{This work} & \multicolumn{2}{c}{MD$^b$}}
\startdata
$      B$       &        19929&27860(42)        &       19929&27880      & &    20677&80826(76)          &    20677&80911    \\
$  10^3D$       &           29&06925(96)        &          29&0683       & &       31&29739(82)          &       31&2977     \\
$  10^9H$       &           --5&10(60)          &          --5&14        & &      --6&31(28)             &      --5&74       \\
$ \lambda_B$    &       158236&114(40)          &      158236&063        & &   158244&81(26)             &   158244&75       \\
$\lambda_D$     &            0&283118(39)       &            0&283426    & &        0&29555(26)          &        0&29413    \\
$  \gamma_B$    &        --155&8017(61)         &       --155&7908       & &     --161&660(24)           &    --161&666      \\
$ 10^3\gamma_D$ &          --0&4366(78)         &         --0&4510       & &      --0&4889(85)           &      --0&4855     \\
$       b_F$    &    \multicolumn{2}{c}{--}     &  \multicolumn{2}{c}{--}& &     --39&784(17)            &     --39&783(11)$^c$\\
$       c$      &    \multicolumn{2}{c}{--}     &  \multicolumn{2}{c}{--}& &       95&359(30)            &       95&340(48)$^c$\\
$     eQq$      &    \multicolumn{2}{c}{--}     &  \multicolumn{2}{c}{--}& &      --3&737(78)            &      --3&614(60)$^c$\\
$   10^3C_I$    &    \multicolumn{2}{c}{--}     &  \multicolumn{2}{c}{--}& &      --7&57$^d$             &      --4&4(15)$^c$  \\
\hline
$\sigma^2$      &             0&74              &  \multicolumn{2}{c}{--$^e$}& &         0&54                &  \multicolumn{2}{c}{--} \\
\enddata
\begin{itemize}
\item [$^a$] Uncertainties (1$\sigma$) are in units of the last
significant digit.
\item [$^b$] Martin-Drumel et al. (\citeyear{2015ApJ...799..115M}).
\item [$^c$] Klaus et al. (1996).
\item [$^d$] Kept fixed at the computed value. See text.
\item [$^e$] Spectroscopic parameters are computed values derived from the isotopically invariant fit.
\end{itemize}\label{tab:2}
\end{deluxetable}

\begin{deluxetable}{cr@{.}lr@{.}ll}
\tablecolumns{4}
\tablewidth{0pc}
\tablecaption{Isotopically invariant parameters$^a$ for the SO radical}
\tablehead{\colhead{Parameter}   & \multicolumn{2}{c}{This work}    & \multicolumn{2}{c}{MD} & \colhead{Units}}
\startdata
 $U_{1,0}$$^c$                &   3757&50375(287)     &     3757&49542(195)              &     cm$^{-1}\cdot$amu$^{1/2}$  \\
 $U_{2,0}$$^c$                &    -68&32282(682)     &      -68&32167(622)              &     cm$^{-1}\cdot$amu          \\
 $U_{3,0}$$^c$                &      0&47701(898)     &        0&47556(818)              &     cm$^{-1}\cdot$amu$^{3/2}$  \\
 $U_{4,0}$$^c$                &     -0&04119(509)     &       -0&04041(463)              &     cm$^{-1}\cdot$amu$^{2}$    \\
 $U_{5,0}$$^c$                &      0&00519(104)     &        0&005031(934)             &     cm$^{-1}\cdot$amu$^{5/2}$  \\
 $U_{0,1}$                    & 230416&7278(172)      &   230387&59070(231)              &     MHz$\cdot$amu  \\
 $U_{1,1}$                    &  -6001&5217(148)      &    -6001&6578(164)               &     MHz$\cdot$amu$^{3/2}$\\
 $U_{2,1}$                    &     25&4579(211)      &       25&7585(267)               &     MHz$\cdot$amu$^{2}$    \\
 $U_{3,1}$                    &     -0&5585(120)      &       -0&8254(175)               &     MHz$\cdot$amu$^{5/2}$\\
 $U_{4,1}$                    &     -0&12215(270)     &       -0&01197(508)              &     MHz$\cdot$amu$^{3}$    \\
 $U_{5,1}$                    &     -0&001253(218)    &       -0&022075(633)             &     MHz$\cdot$amu$^{7/2}$  \\
 $U_{0,2}$                    &     -3&8542930(244)   &       -3&8543097(229)            &     MHz$\cdot$amu$^{2}$  \\
 $U_{1,2}$                    &     -0&004668(126)    &       -0&004547(114)             &     MHz$\cdot$amu$^{5/2}$  \\
 $U_{2,2}$                    &     -0&0010726(628)   &       -0&0011636(577)            &     MHz$\cdot$amu$^{3}$    \\
 $U_{0,3}$                    &     -7&8512(835)E-06  &       -7&8489(791)E-06           &     MHz$\cdot$amu$^{3}$  \\
 $U_{0,4}$                    &     -1&023(156)E-09   &       -1&040(146)E-09            &     MHz$\cdot$amu$^{4}$  \\
 $U^{\gamma}_{0,0}$           &  -1787&3375(287)      &    -1787&4277(276)               &     MHz$\cdot$amu  \\
 $U^{\gamma}_{1,0}$           &    -45&818(188)       &      -45&508(176)                &     MHz$\cdot$amu$^{3/2}$  \\
 $U^{\gamma}_{2,0}$           &      2&0778(143)      &        1&871(133)                &     MHz$\cdot$amu$^{2}$    \\
 $U^{\gamma}_{3,0}$           &      0&0721(227)      &        0&0986(216)               &     MHz$\cdot$amu$^{5/2}$  \\
 $U^{\gamma}_{0,1}$           &     -0&060730(198)    &       -0&060460(184)             &     MHz$\cdot$amu$^{2}$  \\
 $U^{\gamma}_{1,1}$           &      0&00454(117)     &        0&00439(108)              &     MHz$\cdot$amu$^{5/2}$  \\
 $U^{\lambda}_{0,0}$          & 157785&078(303)       &   157795&4203(201)               &     MHz   \\
 $U^{\lambda}_{1,0}$          &   2978&880(153)       &     2979&597(136)                &     MHz$\cdot$amu$^{1/2}$  \\
 $U^{\lambda}_{2,0}$          &    114&524(188)       &      112&732(159)                &     MHz$\cdot$amu           \\
 $U^{\lambda}_{3,0}$          &     11&905(107)       &       13&6225(953)               &     MHz$\cdot$amu$^{3/2}$\\    
 $U^{\lambda}_{4,0}$          &      0&6547(290)      &       -0&1057(327)               &     MHz$\cdot$amu$^{2}$    \\  
 $U^{\lambda}_{5,0}$          &      0&05997(282)     &        0&22948(606)              &     MHz$\cdot$amu$^{5/2}$\\    
 $U^{\lambda}_{6,0}$          &     -0&001964(177)    &       -0&018122(580)             &     MHz$\cdot$amu$^{3}$    \\  
 $U^{\lambda}_{0,1}$          &      3&24178(138)     &        3&24222(125)              &     MHz$\cdot$amu \\  
 $U^{\lambda}_{1,1}$          &      0&14168(874)     &        0&14226(791)              &     MHz$\cdot$amu$^{3/2}$  \\
 $U^{\lambda}_{2,1}$          &      0&03660(511)     &        0&03545(466)              &     MHz$\cdot$amu$^{2}$    \\
 $U^{\lambda}_{0,2}$          &      5&626(387)E-05   &        5&374(357)E-05            &     MHz$\cdot$amu$^{2}$  \\
 $U^{b_f}_{0,0}$              &    -39&7904(214)      &      -39&7830(110)                 &     MHz  \\
 $U^c_{0,0}$                  &     95&3590(315)      &       95&3400(480)                 &     MHz  \\
 $1.5\times U_{eQq}$            &     -3&5687(859)    &       -3&6140(600)                 &     MHz  \\
 $U^{C_I}\times \mu^{-1.0}$   &     -0&00757          &       -0&00440(150)                &     MHz  \\
\hline
 $\sigma^2$                   & \multicolumn{2}{c}{0.7491}     & \multicolumn{2}{c}{0.8997} &                     \\
\enddata
\begin{itemize}
\item [$^a$] Uncertainties (1$\sigma$) are in units of the last significant digit.
\item [$^b$] Martin-Drumel et al. (\citeyear{2015ApJ...799..115M}).
\item [$^c$] The units for these constants have been corrected: in MD these parameters were missing
the amu dependencies. 
\end{itemize}
\label{tab:3}


\end{deluxetable}

\begin{deluxetable}{lr@{.}lr@{.}lr@{.}l}
\tablecolumns{6} 
\tablewidth{0pc}
\tablecaption{Born-Oppenheimer Breakdown coefficients$^a$ derived from the fit}
\tablehead{\colhead{Parameter} & \multicolumn{2}{c}{This work} & \multicolumn{2}{c}{MD$^b$} & \multicolumn{2}{c}{Klaus et al.$^c$}}
\startdata
$\Delta^{S}_{0,1}$            &      -1&9737(16)        &       -1&9725(14)   &       -1&9772(58)   \\
$\Delta^{O}_{0,1}$            &      -2&7015(20)        &       -2&7175(57)   &       -2&7247(34)   \\
$\Delta^{S}_{1,0}$            &       0&409(35)         &        0&395(32)    &        \multicolumn{2}{c}{\nodata}    \\
$\Delta^{O}_{1,0}$            &      -0&2664(89)        &        0&1338(40)   &        \multicolumn{2}{c}{\nodata}     \\
$\Delta^{\lambda,S}_{0,0}$    &       0&866(55)         &       -0&110(85)    &        0&211(89)  \\
$\Delta^{\lambda,O}_{0,0}$    &       1&494(46)         &        1&427(87)    &        1&330(48)  \\
\enddata
\begin{itemize}
\item [$^a$] Uncertainties (1$\sigma$) are in units of the last
significant digit.
\item [$^b$] \citet{2015ApJ...799..115M}.
\item [$^c$] \citet{1996JMoSp.180..197K}.
\end{itemize}
\label{tab:4}
\end{deluxetable}

\end{document}